\documentclass{iopart}
\usepackage{iopams}

\newcommand{\gt}{{\tilde g}}

\begin{document}

\title{Dynamics of a self gravitating light-like shell
with  spherical symmetry}

\author{Jerzy Kijowski$^1$ and Ewa Czuchry$^2$}

\address{$^1$Centrum Fizyki Teoretycznej PAN, Al.\ Lotnik\'ow,
32/46, 02-668 Warszawa, Poland}
\ead{kijowski@cft.edu.pl}

\address{
$^2$Instytut
Problem\'ow J\c{a}drowych, ul. Ho\.za 69, 00-681 Warszawa, Poland}
\ead{eczuchry@fuw.edu.pl}

\begin{abstract}
A novel Hamiltonian description of the dynamics of a spherically symmetric, light-like, self-gravitating shell is presented. It is obtained via the systematic reduction of the phase space with respect to the Gauss-Codazzi constraints,  {\em model} and rare  procedure  in the Canonical Gravity. The Hamiltonian of the system (numerically equal to the value of the ADM mass) is explicitly calculated in terms of the gauge-invariant ``true degrees of freedom'', i.e. as a function on the reduced phase space. A geometric interpretation of the momentum canonically conjugate to the shell's radius is given. Models of matter compatible with the shell dynamics are found. A transformation between the different time parameterizations of the shell is calculated. The presented model may become a new toy model of quantum gravity.
\end{abstract}

\pacs{04.20.Fy, 04.40.-b. 04.60.Ds}
\submitto{\CQG}

\maketitle

\section{Introduction}

Thin shells of matter play an important role in many fields of
gravitational physics. They are used as building blocks in a wide
range of cosmological models \cite{trr1}-\cite{bws} and provide a
convenient laboratory for testing properties of interaction
between gravitation and matter. Recently, shells of light-like
matter have been used to construct a toy model of quantum gravity
\cite{hs}. In particular, the quantum version of the gravitational
collapse has been analyzed in this way. Most of these models possess
either spherical or cylindrical symmetry, which enables us to
analytically solve the dynamical equations of the system. In the quantum
version of the theory, the spherical symmetry ensures the existence of
a well defined ``position'' (configuration) operator which leads
to a simple description of asymptotic states \cite{hc1}-\cite{hc3}.

The complete ``quantization'' of these gravitational toy models
cannot be obtained unless we are able to decipher the underlying
canonical structure of the corresponding phase space and to find
the Hamiltonian function generating its dynamics as a function of
appropriately chosen canonical variables. Such a reduction of the theory -- an elimination of gauge degrees of freedom and a formulation of the dynamics in terms of the ``true degrees of freedom'' -- was proposed  e.g. by J.A. Wheeler  and B. deWitt. 

The research on a thin matter shell was started by 
Werner Israel. In his seminal papers \cite{shell, shella}  he considered the dynamics of a self-gravitating shell of matter with the aim to find a simple model of gravitational collapse.  The dynamics of such an  object  reduces to a proper tailoring of the two different spacetimes (so called Israel's junction conditions), describing the two sides of the shell. The classical equations of motion of  a  shell carrying  self-gravitating light-like matter  were derived by Barrab\`es and Israel twenty years later \cite{IB}.

However, in many investigations -- like creating a toy model for quantum gravity -- a variation principle or a proper Hamiltonian is needed. Many approaches to this problem 
consisted in guessing a Hamiltonian or a super-Hamiltonian of the system from its equations of motion (e.g. \cite{752aa,752a,752b}). Precisely, in \cite{752aa,752a} Berezin {\it et. al.} studied a quantum black hole modeled by a  thin shell of dust matter. In this paper the equations of motion for such a shell  were calculated from Israel's junction condition and the Hamiltonian of the system was simply set to the derived total mass (energy) of the system. In \cite{752b} H\'{a}j\'{\i}\v{c}ek  investigated the quantum gravitational collapse in a simple case of a spherically symmetric thin shell of dust matter with a fixed rest mass, interacting with its own gravitational field. Here the Hamiltonian of the system was just guessed in a form that  reproduced the equations of motion for the shell.

Nonetheless, equations of motion do not uniquely determine  an action or a Hamiltonian they are generated by (the so called inverse problem of variational principle). Some attempts to derive a Hamiltonian structure in a spherically symmetric case were made (\cite{KWLa,KWLb,louko}), yet they based on an intermediate variational principle. Kraus and Wilczek \cite{KWLa} studied the back reaction in the Hawking effect,  i.e. the self-gravitational interaction of the radiation and 
its interaction with the hole. For this purpose a complete dynamical description of a self-gravitating particle was needed, which was implied by a Hamiltonian derived from a  variational principle guessed for a spherically symmetric system of dust shell and its gravitational field. Friedman, Louko {\it et. al.} analyzed the Hamiltonian dynamics of a massive \cite{KWLb} and a null \cite{louko} spherical dust shell. Their Hamiltonian was, similarly as in \cite{KWLa}, obtained from a postulated variational principle.  In a massive case, with the interior mass fixed,  Friedman and Louko's Hamiltonian reduces to the one found by Kraus and Wilczek in the limit of a massless shell. Unfortunately the results were very complicated and gauge dependent Hamiltonians -- functions of non local momenta -- were not useful to serve as a framework for quantization. 

The turn in research on deriving dynamics of thin matter shells was made by H\'{a}j\'{\i}\v{c}ek  (\cite{h1}, \cite{shell1}) who looked for a simple "toy model" for quantum gravity and quantum collapse. In \cite{h1} the dynamics of the spherically symmetric shell is obtained from a super-Hamiltonian on which a set of conditions was imposed. The work \cite{shell1} by  H\'{a}j\'{\i}\v{c}ek and one of us (J.K.) contains derivation of both the super Hamiltonian and the symplectic structure  for discontinuous fluid. It was obtained  from the Einstein-Hilbert variational principle for  ideal fluid, without assuming any additional symmetries. 

 We have done a considerable simplification of this theory by applying the theory of distributions (in \cite{JKC1} for a lightlike shell and \cite{KC} for a massive shell).  The work \cite{JKC1} on the dynamics of a self-gravitating null shell of matter is the first one   using fully gauge-invariant, intrinsic geometric objects
encoding physical properties of both the shell  and the light-like
matter living on the shell. Moreover, null matter was treated in  a fully dynamical (and not phenomenological)
way. All its  properties  were encoded
in a matter Lagrangian, without assuming any of its additional properties (like being dust or ideal fluid). In the present paper  we  apply these results  to the case of a spherically symmetric shell of null matter.  The results obtained here provide a successful implementation of the programme started by Wheeler and deWitt, continued by H\'{a}j\'{\i}\v{c}ek,  and can be used as a simple ``toy model'' for quantum gravity.

In general, a matter shell may be considered as a singularity of
space-time geometry arising when two different, smooth
four-geometries are tailored together along a 3-dimensional hypersurface $S$.
In this paper we consider the case when $S$ is a null
(``light-like'') surface. Here, {``tailoring''} means that the
(properly understood) internal three-geometries of $S$, induced
from each of the two four-geometries, do coincide (\cite{IB,JKC1,H-Kou1,H-Kou2,H-Kou3}). The four-dimensional connection
coefficients $\Gamma^\lambda_{\mu\nu}$ may, however, be
discontinuous across $S$. The curvature tensor density of such 
spacetime contains derivatives of those discontinuities and,
whence, is singular. Such   singularity may be nicely described in
the sense of distributions as ${\cal G}^a_{\ b} = {\bf G}^a_{\ b}
{\boldsymbol\delta}_S$, where ${\boldsymbol\delta}_S$ is the Dirac
delta distribution concentrated on $S$ and ${\bf G}^a_{\ b}$ is a
smooth tensor-density field living on $S$. Contrary to the
time-like shell, where the tensor-density ${\bf G}^a_{\ b}$ may be
decomposed into a product of the 3-volume element on $S$ and the
tensor $ G^a_{\ b}$, here such a decomposition is meaningless
because of the degeneracy of the 3-metric. Hence, the
``energy-momentum tensor'' $G$ cannot be defined. Nevertheless, we
have shown in  \cite{JKC1} how to construct the
``energy-momentum tensor-density'' ${\bf G}$ in a nice, geometric
way. Using this quantity, the dynamics of the composed
``null-matter-shell + gravity" system was derived from first
principles, in both the Hamiltonian and the Lagrangian approach,
and for  general null matter.

For this purpose an appropriate Lagrangian $L = L_{\textrm{grav}}
+L_{\textrm{matter}}$ was used, where $L_{\textrm{grav}}=\frac
1{16 \pi} \sqrt{|g|} \ {R}$. Here ${R} = R_{\textrm{reg}} +
R_{\textrm{sing}}$ is the 4-dimensional curvature scalar,
$R_{\textrm{reg}}$ denotes its regular part and $\sqrt{|g|}
R_{\textrm{sing}}=- \mbox{\rm sing}({\cal G}) = - {\bf G}^{\mu\nu}
g_{\mu\nu} {\boldsymbol\delta}_S$ is its singular part.

Assuming the spherical symmetry of the shell, we obtain  a
simple Hamiltonian system in this way. In the case of a massive shell its complete
description was given in \cite{shell1a} and \cite{spher-lagr1,spher-lagr2}.
Here, we are going to present similar results for the case of a
null shell. Similarly as was done in \cite{shell1a}, a spherical
symmetry condition is imposed at the Hamiltonian level only. (On
the other hand, one could also impose this symmetry already at the
level of variational (Lagrangian) formulation, as was done in
\cite{spher-lagr1,spher-lagr2}.) As a result, we obtain the Hamiltonian dynamics of
the system described in terms of gauge-independent variables and
their conjugate momenta.

The fundamental difference between the null and the massive case
consists in the fact, that there are no massive shells without
matter. On the other hand, gravitational ``shock waves'', i.e.
null shells without any matter, are perfectly allowed by the
theory. Nevertheless, our theory admits also null-like matter,
which couples consistently to gravity. We have analyzed several
models of such matter. In most cases it couples to an isolated
horizon only, which is rather a trivial case. However, other interesting
and self-consistent models have been found (see 
\ref{lagrangians}). They lead to non trivial dynamical systems.

The space of initial data of the ``matter + gravity'' system can
be parameterized by the following space of functions:
\[
{\cal P} := \{ ( g_{kl}, P^{kl}, z^K , p_K )  \} \ ,
\]
where  $g_{kl}$ is a three-dimensional metric on a spacelike
Cauchy surface ${\cal C}$, and   $P^{kl}$ are appropriate ADM
(\cite{ADM}) momenta describing external curvature of this
surface. Moreover, $z^K$ describe configuration variables of the
matter and $p_K$ describe their conjugate momenta. We limit ourselves to
the topologically trivial case ${\cal C}\simeq {\bf R}^3$ and
assume that the geometry  of ${\cal C}$ is asymptotically flat at
infinity.

The above phase space is equipped with the canonical
pre-symplectic structure:
\begin{equation} \label{Omega-gen-male-0}
\Omega :=\frac 1{16 \pi} \int_{\cal C} \left\{ \delta P^{kl} (x)
\wedge \delta g_{kl} (x) \right\} d^3x  + \int_{\cal
C} \left\{ \delta p_K(x) \wedge\delta z^K(x)
  \right\} d^3 x\ .
\end{equation}
Assuming that the matter is concentrated on a shell two-surface $S_t=S
\cap {\cal C} \subset {\cal C}$, we obtain:
\begin{equation} \label{Omega-gen-male}
\Omega := \frac 1{16 \pi} \int_{\cal C} \left\{ \delta P^{kl} (x)
\wedge \delta g_{kl} (x) \right\} d^3x + \int_ {S_t}
\left\{ \delta p_K(x) \wedge\delta z^K(x)
  \right\} d^2 x\ .
\end{equation}

The structure of the paper is following. In Section \ref{symmetry}
we analyze the above pre-symplectic form and reduce it with
respect to both the spherical symmetry and the Gauss-Codazzi
constraints fulfilled by the ADM data. It turns out that the radial
component of the constraint equations gives us an ``equation of
state'' which has been previously postulated in 
\cite{louko}. The exact solution of the constraint equations is
obtained in Section \ref{constraints}. Using these results, both
the gravitational part (Section \ref{grav-reduction}) and the
matter part (Section \ref{mat-reduction}) of the symplectic form
are reduced. Then, an exact form of the Hamiltonian of the system
is derived which encodes its equations of motion. These equations
are solved with respect to the Schwarzschild time in Section
\ref{dynamics}. Next, we perform a transition from the Schwarzschild to
the Minkowski time (Section \ref{time}), and solve the resulting
equations of motion. In  \ref{lagrangians} we discuss
different models of null matter.

This paper is a continuation of  \cite{JKC1} and
\cite{JKC2}. Most of the  results presented here are contained in
the Ph.D. thesis \cite{thesis} of one of us (E.C.).

\section{Spherical symmetry} \label{symmetry}

The spherical symmetry of the system implies the existence of
spherical coordinates $(x^1,x^2,x^3)=(\theta , \varphi, r)$ on the
Cauchy surface ${\cal C}$, so that initial data $( g_{kl},
P^{kl}, z^K , p_K )$ are invariant with respect to rotations. For
calculational purposes, we may choose the radial coordinate in
such a way that it is constant on the shell (i.e. the history of
the shell is given by the equation $r=x^3\equiv \zeta ={\rm const}$).
This is {\em not} a physical assumption, but merely a choice of a
gauge, as legal as any other gauge fixing conditions (e.g.
conditions used in  \cite{shell1,shell1b,shell1a}). It considerably
simplifies  our calculations (e.g. we have
${\boldsymbol\delta}_S= \Delta(x^3)$ in this coordinate system,
where  $\Delta$ denotes the one-dimensional Dirac delta).

Another gauge condition which we use in this paper is the
continuity of all the 10 components of the metric. In many
approaches only the internal geometry of the shell is supposed to
be continuous across the shell. Our condition is stronger. Again,
this is {\em not} a physical restriction imposed on the theory but
merely a gauge fixing condition which allows us to use the theory of
distributions to calculate the (singular!) curvature tensor. In
particular, our condition implies an equivalence of the affine
structure of null geodesics on the shell, as seen from both sides
of the shell. Without our gauge condition, this equivalence must
be imposed separately (see \cite{spher-lagr1,spher-lagr2} for further
discussion of the role of this gauge condition).

Due to spherical symmetry the two four-geometries, which we
tailor together across the shell must be Schwarzschild. Moreover,
due to the topological triviality of the Cauchy surface ${\cal C}$,
the internal portion of the spacetime must be Minkowski.

We assume that asymptotically, for $r \rightarrow \infty$ and $r
\rightarrow 0$, the radial coordinate is ``asymptotically flat'',
i.e. equals to the Schwarzschild radius outside and the Minkowski
radius inside the shell.

Any spherically symmetric three-metric on ${\cal C}$ has the
following form:
\begin{equation}
{g}_{kl}=\left[
\begin{array}{ccc} l(r) \ \gamma_{AB}
 & \vline &0\\ & \vline & \\
\hline & \vline & \\ 0  & \vline & n^2 (r)
\end{array}
\right] \ ,
\end{equation}
where  $l$ and $n$ are  the functions of a radial coordinate
 $r$, and $\gamma_{AB}$ (where $A,B=1,2$ label angular coordinates)
denotes the standard metric on the unit sphere:
\begin{equation}
\gamma_{AB} = \left( \begin{array}{cc} {1} & {0} \\ {0} & {\sin^2
\theta }
\end{array} \right) \ .
\end{equation}
Spherical symmetry also implies  that the components $P^{kl}$ of
ADM momenta assume the following form
\begin{eqnarray}
P^{AB} & = & \frac 12 u (r) \ \gamma^{AB} \sqrt{\det \gamma }  \ ,
\nonumber \\ P^{33} & = & \frac {f(r)}{n(r)} \ \sqrt{\det \gamma }
\ , \nonumber \\ P^{3A} & = & 0 \ , \label{ADMmomenta}
\end{eqnarray}
where $u$ and $f$ are again the functions of the radial coordinate.
These functions are piecewise smooth outside of the sphere $r =
\zeta$, whereas the metric coefficients $n$ and $l$ are supposed
to be also continuous  at $r = \zeta$.

The Einstein equations imply constraints which must be satisfied by
the above data, namely the Gauss-Codazzi equations for the
components ${\cal G}^0{_\mu}$ of the Einstein tensor density.
The standard decomposition of ${\cal G}^0{_\mu}$ into the spatial
(tangent to ${\cal C}$) part and the time-like (normal to ${\cal C}$)
part gives us respectively:
\begin{eqnarray}
{\cal G}^0{_l} &=&- P_l{^k}{_{|k}}\ , \label{GCww} \\
 2{\cal G}^0{_\mu} n^{\mu} &=&
- sg\stackrel{(3)}{R}+s\left(P^{kl}P_{kl}-\frac{1}{2}\label{GCws}
P^2\right)\frac1{g}  \, ,
\end{eqnarray}
 where $n$ is the future-oriented vector, orthonormal
to the Cauchy surface ${\cal C}$ and $s$ denotes the sign of $g^{30}$.
In  \cite{JKC} (see also \cite{JKC1}) we have shown how to
decompose these equations into the regular and the singular
(proportional to Dirac delta distribution on $S_t={\cal C}\cap S$)
parts. The regular part of the vector constraint reads:
\begin{equation}
  \label{rwd}{\rm reg}\left( P_l{^k}{_{|k}}\right)=0 \ ,
\end{equation}
whereas the regular part of the scalar constraint reduces to:
\begin{equation}
  \label{rsd} {\rm reg} \left(
 g \stackrel{(3)}{R} - \left(P^{kl}P_{kl}-\frac{1}{2}
 P^2\right)\frac 1g  \right)= 0 \, .
\end{equation}
The singular part of the divergence of the ADM~momentum $P_{kl}$
contains derivatives in the direction of $x^3$ :
\[
  {\rm sing}(P_l{^k}_{|k}) ={\rm sing}(\partial_3 P_l{^3})=
  \boldsymbol\Delta(x^3)[P_l{^3}]\ ,
\]
where the square brackets $[]$ denote the jump of $P_l{^3}$ between the two
sides of the singular surface. Hence, the right-hand-side of
(ref{GCww}) has the form
\begin{equation}\label{wwd}
  P_l{^k}_{|k} = [P_l{^3}] \boldsymbol\Delta(x^3)\ .
\end{equation}
 In  \cite{JKC1} we have proved
that the components $P^{kl}$ are regular, whereas the singular part of
the three-dimensional curvature scalar reduces to the jump of the
external curvature $k$ of the two-dimensional surface $S_t \subset
{\cal C}$ across the shell:
\[ \textrm{sing}(g\stackrel{(3)}{R})= 2 g
{\sqrt{\gt^{33}}} [k] \boldsymbol\Delta(x^3) = 2 \sqrt{\gamma l}[
k] \boldsymbol\Delta(x^3)\, ,
\]
Hence, the right-hand-side of (\ref{GCws}) has the form:
\begin{equation}\label{c}
sg\stackrel{(3)}{R} -s \left(P^{kl}P_{kl}-\frac{1}{2}
P^2\right)\frac1{g}
  = 2  \sqrt{\gamma l}[k] \boldsymbol\Delta(x^3) \, .
\end{equation}
Due to the Einstein equations, these objects must match the matter
energy-momentum tensor. {\em A priori}, there are serious
difficulties concerning the definition of such an object in case of the
matter living exclusively on the null surface $S$. We have shown
in  \cite{JKC1} that such a quantity may be consistently
defined. It is a three-dimensional {\em tensor-density} denoted by
$\tau^a{_b}$ (one three-dimensional index $a=0,1,2$ up and one
index down!). We stress that, due to the degeneracy ($\sqrt{\det
g_{ab}}=0$) of the metric tensor $g_{ab}$ on $S$, it is impossible
to find any {\em tensor} representation of this quantity. Also,
the otherwise trivial ``rising of indices'' is forbidden,
whereas the ``lowering'' becomes a non-invertible (losing
information) procedure!

Naively, one could expect Gauss-Codazzi constraints in the form:
\[
  {\cal G}^0{_\mu} = 8\pi\tau^0{_\mu}\ .
\]
We stress, however, that there is no way to define the right-hand
side as components of any well defined four-dimensional object! In
our notation, where $x^3$ is constant on $S$, only $\tau^0{_b}$
makes sense! Fortunately, we have shown in \cite{JKC1} (see also
\cite{JKC}) that, due to the null character of $S$, the singular
part of the constraint equations contains only three conditions.
Indeed, the orthogonal (to $S$) part of the constraint equation
coincides with one of the tangent parts of these equations, namely
the component along the null vector on $S$. (The fourth
constraint, existing in a non-degenerate case, is replaced here by
the degeneracy condition ``$\det g_{ab}=0$'' for the metric on
$S$.) The tangent (to $S$) part of ${\cal G}^0{_b}$ splits into
the two-dimensional part tangent to $S_t$ and the transversal part
(along null rays). The first one gives us:
\begin{equation}
  \label{PBt} \left[ P^3{_B}\right]=-8\pi\tau^0{_B}\ .
\end{equation}
In the spherically symmetric case both left- and right-hand-side must
vanish and the above constraint is fulfilled automatically. The
remaining null tangent part of the Einstein equations reduces, as we
have shown in \cite{JKC1}, to the following constraint:
\begin{equation}
 \label{plk}
 \left[\frac{P^{33}}{\sqrt{\gt^{33}}}+ \sqrt{\det g_{AB}} k \right]=0\ .
\end{equation}
In our notation it can be rewritten in the form which has been
postulated in  \cite{louko}:
\begin{equation}
\label{eqos}
 -s\frac{p}{n}+e = 0\ ,
\end{equation}
where we have denoted:
\begin{eqnarray}
  p:&=&\, \frac1{8\pi}\textrm{sing}\, \left({P^{33}}_{|3}\right)
  = \frac1{8\pi} [ P^{33} ]\label{p-def} \ ,\\
  e:&=&\, \frac1{16\pi} \textrm{sing}\, \left(g\stackrel{(3)}{R} -
  (P^{kl}P_{kl}-\frac{1}{2} P^2)\frac1{g}\right)
  = \frac1{8\pi} \sqrt{\gamma l} [k] \ .\label{e-def}
\end{eqnarray}
The ``equation of state'' (\ref{eqos}) is a consequence of the
``nullness'' of matter and does not come from its internal
properties.

The equations (\ref{PBt}), (\ref{plk}), together with (\ref{rwd}) and
(\ref{rsd}), provide the complete set of constraints fulfilled by
the initial data $(l(r),n(r),u(r),f(r))$ of the theory.

To reduce the phase space with respect to these constraints we
will proceed as in  \cite{shell1a}, but with some
modification. The equation (\ref{p-def}), together with the vector
constraint (\ref{rwd}), can be written in terms  of momenta
(\ref{ADMmomenta}):
\begin{equation} \label{vector-spherical}
- 8\pi s\frac {p}n \ \Delta (r - \zeta ) = f^\prime - \frac 12
\frac un \ l^\prime \ ,
\end{equation}
where prime  denotes the radial derivative $\partial /\partial r$
and $\Delta$  is the one-dimensional Dirac delta. The equation
(\ref{e-def}), together with (\ref{rsd}), may be rewritten in the
analogous way:
\begin{equation} \label{scalar-spherical}
- 8\pi e (p,l,n) \ \Delta (r - \zeta ) = \left( \frac {l^\prime}n
\right)^\prime - n - \frac 14 \frac {(l^\prime )^2}{ln} + \frac 14
\frac nl \ f^2 - \frac 12 fu \ .
\end{equation}

The above constraints generate two dimensional group of space-time
reparameterizations, where the variables  $(t,r)$ may be replaced by the
new variables  $(\tilde{t} , \tilde{r} )$, preserving the spherical
symmetry of our system. Gauge transformations enter as the degeneracy
directions  of the symplectic structure ${\cal P}_{sym}$, obtained by the
restriction of the form $\Omega$ from ${\cal P}$ to the space of the
spherically symmetric data which we denote by ${\cal P}_{sym}$. In
order to calculate this restriction let us consider 
(\ref{Omega-gen-male}) in our specific case and integrate over
angular variables $x^1$ and $x^2$. Hence we obtain the following
symmetric structure in ${\cal P}_{sym}$:
\begin{equation} \label{Omega}
\Omega = \int_0^\infty \left( \frac 14 \delta u(r) \wedge \delta
l(r) + \frac 12 \delta f(r) \wedge \delta n(r) \right) dr + 4\pi
  \delta ( P_K l(\zeta) )\wedge\delta z^K\ .
\end{equation}
where we have denoted $p_{K}=:\sqrt{\det g_{AB}} P_{K}$. Moreover, for
purely calculational reasons it is convenient to write the 2-form
 $\Omega$ as an exterior derivative  $\Omega = \delta \Theta$ of the
 following 1-form:
\begin{equation} \label{Lambda1}
\Theta= \int_0^\infty \left( \frac 14  u(r) \delta l(r) - \frac 12
n(r) \delta f(r) \right) dr + 4\pi l(\zeta)  P_K
\delta z^K\ .
\end{equation}

As already mentioned, the Cauchy surface  ${\cal C}$ is tailored from
two pieces: a piece of a 3-surface in the flat Minkowski spacetime
for $r < \zeta$ and a piece of a 3-surface in the Schwarzschild
space-time for $r>\zeta$.

\section{Solution of the constraint equations} \label{constraints}

In order to solve the constraint equation we have to impose a gauge
condition which enables us to  uniquely  fix the time coordinate.
For purely technical reasons we start with the family of
$\beta$-gauge conditions proposed in  \cite{Jez-Kij} in order
to prove the positivity of the ADM mass. When written
in the spherical symmetric case, the condition reduces to: $
\beta P^{33} g_{33} + P^{AB} g_{AB} = 0$, where $\beta $ is a
fixed constant. Hence, we have:
\begin{equation} \label{beta}
\frac un =  - \beta \frac fl \ .
\end{equation}
Assume that $\beta < -1 $. Inserting the above relation to the
vector constraint (\ref{vector-spherical}) we see that, outside
the shell, our initial data must fulfill the following equation
\begin{equation}
f^\prime + \frac 12 \beta \ \frac {l^\prime}l f = 0  \ .
\end{equation}
This implies that the function    $\log (f \ l^{\frac {\beta}2} )
$ is constant everywhere outside of the shell. Hence, there are
two constants $A_+$ and $A_-$ such that:
\begin{equation} \label{f}
f = \left\{ \begin{array}{ll} A_+ l^{- \frac {\beta}2 } &
\mbox{for} \ \  r > \zeta \ , \\ A_- l^{- \frac {\beta}2 } &
\mbox{for} \ \  r < \zeta \ .
\end{array} \right.
\end{equation}
The jump  $(A_+ - A_- )$ is determined by the constraint equation
(\ref{vector-spherical}) -- i.e. the only singular term on the
right hand side of that equation is produced by the jump of $f$
and is equal to  $(A_+ - A_- )(l(\zeta))^{- \frac {\beta}2 }
\Delta(r-\zeta ) $. Denote the {\em physical}  radius of the shell
by $R$:
\[
R := \sqrt{l(\zeta)}\ ,
\]
and the normalized radial component of the momentum by $U$:
\begin{equation} \label{U}
U := \frac {4 \pi s p}{n(\zeta ) \sqrt{l(\zeta )}} = \frac {4 \pi s
p}{n(\zeta ) R }\ .
\end{equation}
The equation  (\ref{vector-spherical}) implies:
\begin{equation}
  \label{A_}
  A_+ - A_- = - 2 U R^{1 + \beta } \ .
\end{equation}
To completely solve the constraint equation, the boundary conditions
must be imposed, i.e. for $r \rightarrow \infty$ the initial data must
be asymptotically flat, and for  $r \rightarrow 0$ must be
regular. This implies that  $l$ must behave like $r^2$ both at
zero and  at infinity, whereas $f$ must vanish at infinity.
Consequently, for  $\beta < -1$, we obtain $A_+ = 0$:
\begin{equation} \label{f-}
f = \left\{ \begin{array}{ll} 0 & \mbox{for} \ \  r > \zeta \ , \\
2 U R^{1 + \beta } l^{ - \frac {\beta}2 } & \mbox{for} \ \ r <
\zeta \ .
\end{array} \right.
\end{equation}

Now, we are going to solve the scalar constraint equation
(\ref{scalar-spherical}). Using the gauge condition (\ref{beta})
and the vector constraint (\ref{f}), we obtain the following
equation which is valid outside and inside of the shell:
\begin{equation} \label{scalar-0}
0 = \left( \frac {l^\prime}n \right)^\prime - n - \frac 14 \frac
{(l^\prime )^2}{ln} + \frac {1+2\beta }{4 l^{1+\beta} } n A_\pm^2
\ .
\end{equation}
Consider the following quantity:
\begin{equation}
k := \frac {l^\prime}{n \sqrt[4]{l}} \ .
\end{equation}
We see that (\ref{scalar-0}) may be rewritten as follows
\begin{equation}
0 = \frac {\sqrt[4]{l}}{n} \ k^\prime - 1 + \frac {1+2\beta }{4
l^{1+\beta} }  A_\pm^2  \ ,
\end{equation}
or, equivalently,
\begin{equation}
0 = \left\{ k^2 - 4 \sqrt{l} \left( 1 + \frac {A_\pm^2}{4
l^{1+\beta }} \right) \right\}^\prime \ .
\end{equation}
This means that the function in the brackets must be piecewise
constant. The regularity of the metric at $r = 0$ implies that for
a small $r$ the function  $l$ shall behave like   $n^2 r^2$. This
implies  that $k^2$ behaves like  $4 \sqrt{l}$ and, consequently,
the corresponding {\em internal} constant vanishes. Denoting the
remaining {\em external} constant by ,,$-8 H$'', we have, due to
$A_+=0$:
\begin{equation} \label{l-prime}
\frac {l^\prime}n = \left\{ \begin{array}{ll}
  \pm  2 \sqrt{l} \sqrt{1 - \frac {2H}{\sqrt{l}} } & \mbox{for} \ \ r > \zeta \ , \\
2 \sqrt{l} \sqrt{1  + \frac {A_-^2}{4  l^{1+\beta }} }   &
\mbox{for} \ \  r < \zeta \ ,
\end{array} \right.
\end{equation}
where the sign ``$\pm$'' appearing in the first line depends upon
the sign of  $l^\prime$. Outside of the shell this sign may change
at the points where the expression under the square root vanishes. On the
contrary, inside of the shell this sign is always positive,
because for a space-like hypersurface in Minkowski's space-time
$l$ is always increasing.

The value of $H$ is equal to the ADM energy calculated at $r
\rightarrow \infty$. This is implied by the fact that, for large
values of $r$, we have  $l(r) \sim r^2$. Consequently, due to
(\ref{l-prime}), we obtain:
\begin{equation}
g_{33} = n^2 \sim \frac 1{1 - \frac {2H}{r} } \ .
\end{equation}
We expect that the ADM mass $H$ will play the role of the Hamiltonian
of the total ``gravity + matter'' system (cf. \cite{K1,pieszy11,pieszy12}).  The numerical value of $H$
may be obtained from the singular part of
(\ref{scalar-spherical}). Since the singular part of the right
hand side of this equation is equal to the jump $\frac
{l^\prime}n$, we obtain:
\begin{equation}
- 8 \pi e = 2 R \left( \epsilon  \sqrt{1 - \frac {2H}{R} } -
\sqrt{1  + \frac {A_-^2}{4 R^{2+2\beta }} }  \  \right) \ ,
\end{equation}
where  $\epsilon$ denotes the sign of    $l^\prime$ outside the
shell. Now we insert the value of $A_-$ calculated from (\ref{A_})
(where $A_+=0$), our ''equation of state'' (\ref{eqos}) and the
value of $p$ calculated from (\ref{U}). In this way we obtain:
\begin{equation} \label{H-aux}
-  U   = \epsilon \sqrt{1 - \frac {2H}{R} } - \sqrt{1 + U^2 } \ ,
\end{equation}
and, consequently:
\begin{equation} \label{H}
H ( R, U ) = \frac R2 \left\{ 1 - \left( \sqrt{1 + U^2  } - U \
\right)^2 \right\} \ .
\end{equation}
The value of   $\epsilon$ can be obtained from 
(\ref{H-aux}):
\begin{equation} \label{eps}
\epsilon = \mbox{sgn}   \left( \sqrt{1 + U^2  } - U \  \right)= 1\
.
\end{equation}

In the next Section we prove that the reduced phase space for the
gravitational degrees of freedom $\tilde{\cal P}$ may be
parameterized globally by the two variables $R$ and $U$, whereas
the function $l$ and the constants $\zeta$, $\beta$ play the role of
{\em gauge parameters}. Given the point $(R,U)$ of this space, a
specific choice of the gauge parameters allows us to reconstruct
the entire Cauchy data, with the values of $A_\pm$ and $H$ given
uniquely by  (\ref{f-}) and (\ref{H}). However, gauge
parameters are not completely arbitrary. One condition is obvious:
$l(\zeta ) = R^2$. Moreover,  inside the shell, $l$ must increase
monotonically from 0 to $R^2$. Because the sign ``$\epsilon$''
given by (\ref{eps}) is positive, outside of the shell $l$ must
increase and, due to the asymptotical flatness, must behave like $r^2$
at infinity. Any function $l$ is allowed provided it fulfills
these conditions. The equations (\ref{l-prime}), (\ref{f-}) and
(\ref{beta}) enable us to completely reconstruct  the data
$(n,l,f,u)$. Within our gauge subspace, given by condition
(\ref{beta}), all states may be obtained in this way. In the sequel
we show  that the states of the system obtained for other values
of   $\zeta$ and $l$, are equivalent up to a gauge transformation.

For negative values of  $\beta$ the equation (\ref{f-}) implies that
the entire external curvature vanishes outside of the shell:
$P^{kl} = 0$. Due to the fact that the external portion of
space-time is  Schwarzschild with the mass $H$, the only surfaces
satisfying this condition are standard surfaces $\{ t =
\mbox{const} \}$.  The Cauchy surfaces corresponding to different
negative values of $\beta$ coincide, therefore, outside of the
shell and differ  only inside of the shell.

\section{Canonical structure of the reduced phase space}

The dynamics of our system will be given uniquely by the Hamiltonian
(\ref{H}), if we only know the reduced symplectic structure expressed
in terms of the gravitational variables  $(R,U)$ and, possibly,
additional matter variables.

It is possible that there is no matter at all,  and the matter
Lagrangian is  equal to zero. Yet, the space-time may still
exhibit singularity and our ``null shell'' becomes a shock wave of
pure gravitation. The phase space is then two-dimensional and
the symplectic form consists of the gravitational part only. Its
construction is given below.

\subsection{Reduction of gravitational part of the symplectic
form}\label{grav-reduction}

We restrict the gravitational part of the symplectic form $\Omega$
given by (\ref{Omega}) to the gauge space  (\ref{beta}) and
express it in terms of parameters $(R,U,l,\zeta )$. It turns out
that this form does not depend upon $l$ and $\zeta$. This proves
that the latter are, indeed, gauge variables. For technical
reasons it is easier  to work with the 1-form $\Theta$ given by
 (\ref{Lambda1}), because the restriction to the gauge
space commutates with the exterior derivative of the form.

Formula (\ref{Lambda1}) may be rewritten as follows:
\begin{equation}
\Theta  =  \int_0^\infty
\left(\frac 14 n \left( \frac un + \beta \frac fl \right) \delta l
- \frac 12 n l^{- \frac {\beta }2 } \ \delta \left(
  f l^{ \frac {\beta }2 }  \right)
\right) +4 \pi R^2 P_K \delta z^K\ .
\end{equation}
Due to gauge condition (\ref{beta}), the first term in the above
formula vanishes. The equation  (\ref{f-}) implies:
\begin{equation}
 f l^{ \frac {\beta }2 }  = 2U  R^{1+\beta } \
B(\zeta - r) \ ,
\end{equation}
where  $B$ denotes    the Heaviside function. Then
\begin{equation}
\left\{ \delta \left(   f l^{ \frac {\beta }2 }  \right) \right\}
(r) = 2 B(\zeta - r) \delta (U  R^{1+\beta } )  \ ,
\end{equation}
and, consequently,
\begin{equation}
- \frac 12 \int_0^\infty  n l^{- \frac {\beta }2 } \left\{ \delta
\left(
 f l^{ \frac {\beta }2 }  \right) \right\} (r) dr  =
 -  \left\{ \int_0^{\zeta} \left( n l^{- \frac {\beta }2 }
\right) (r) dr \right\} \delta (U  R^{1+\beta } ) \ ,
\end{equation}
because of the condition $l(\zeta ) = R^2$ and due to the vanishing of
$f$ outside of the shell. Finally, we obtain
\begin{equation}\label{34}
\Theta=
 - w \delta (U R^{1+\beta } ) +
 4 \pi R^2  P_K \delta z^K \  ,
\end{equation}
where   $w$ denotes
\begin{equation}
w := \int_0^{\zeta} \left( n l^{- \frac {\beta }2 } \right) (r) dr
\ .
\end{equation}
Using   (\ref{l-prime}) we obtain:
\begin{equation}
w = \int_0^{\zeta} \frac n{l^\prime} \  l^{- \frac {\beta }2 }
l^\prime dr = \int_0^{\zeta} \frac {l^{- \frac {1 +\beta }2 }}{2
\sqrt{1 + \left\{ U  \left(  \frac {\sqrt{l}}{R}
\right)^{-(1+\beta )} \right\}^2 } } \ l^{\prime} dr \ .
\end{equation}

The function $l$ is monotonic over the interval $[0,\zeta ]$.
Moreover, we have: $-(1+\beta )
> 0$. Hence, we may calculate the integral with respect to the following
variable:
\begin{equation}
\xi := U  \left(  \frac {\sqrt{l}}{R}  \right)^{-(1+\beta )} \ ,
\end{equation}
over the interval $[0, U]$. We have:
\begin{equation}
w = - \frac {1}{1+\beta} \left( U R^{1+\beta} \right)^{\frac
{2}{1+\beta} - 1} \int_0^{U} \frac {\xi^{- \frac 2{1 +\beta } }}{
\sqrt{1 + \xi^2} } \  d\xi \ .
\end{equation}
The last integral may be denoted as  $F( U )$, where  $F$ is the
indefinite  integral. It turns out that the specific form of $F$
will not be needed. The first (gravitational) part of (\ref{34})
reads:
\begin{eqnarray}
&& \frac {1}{1+\beta} F(U) \left( U R^{1+\beta} \right)^{\frac
{2}{1+\beta} - 1} \delta (U R^{1+\beta }) \\&&= \frac 12 F(U) \delta
\left( U^{\frac {2}{1+\beta} } R^2 \right)  \nonumber \\
& &=
\frac 12 \delta \left( F(U)  U^{\frac {2}{1+\beta} } \ R^2 \right)
- \frac 12 \ R^2 U^{\frac {2}{1+\beta} } F^\prime (U) \delta (U)
\nonumber \\
& &=
 \frac 12 \delta \left( F(U) U^{\frac {2}{1+\beta} } \ R^2 \right)
- \frac 12 \ R^2 \ \frac 1{\sqrt{1+ U^2}} \ \delta (U) \label{u}
\end{eqnarray}
The first term in the above formula is a complete (variational)
derivative. Hence, it vanishes under the exterior derivative, when
we calculate the symplectic form: $\Omega = \delta \Theta$. Therefore, the
second term alone gives us an equally good primitive
form $\tilde{\Theta}$ for $\Omega$:
$\delta\Theta=\delta\tilde{\Theta}=\Omega$. Taking into account
that
$$
  \frac 1{\sqrt{1+ U^2}} \ \delta (U)=\delta\textrm{ arsinh} \ U\
,
$$
we obtain the following formula:
\begin{equation}
\tilde{\Theta} = - \frac 12 \rho \ \delta \mu +4 \pi \rho
 P_K \delta z^K\ ,
\end{equation}
where by $\mu$ we denote the momentum canonically conjugate to the
variable $\rho := R^2$:
\begin{equation}\label{mom-can}
    \mu := \textrm{ arsinh} \ U
\end{equation}
Consequently, we have:
\begin{equation}
\Omega = \delta \tilde{\Theta} =
\frac 12 \delta  \mu \wedge \delta \rho +4 \pi
\delta (\rho P_K) \wedge\delta z^K\ .
\label{symplektyczna}
\end{equation}

Finally the Hamiltonian (\ref{H}) may also be expressed in terms
of variables   $(\mu , \rho )$:
\begin{equation} \label{H-reduced}
 H ( \mu , \rho )=
\frac12\sqrt{\rho}\left(1-e^{-2\mu}\right) .
\end{equation}

For the shock wave, when matter degrees of freedom  vanish from
the very beginning, the symplectic form is of the following form
\begin{equation}
\Omega = \frac 12 \delta  \mu \wedge \delta \rho ,
\label{symplektyczna1}
\end{equation}
and the Hamiltonian is given by formula (\ref{H-reduced}), with $\mu$
and $\rho$ being now canonical variables.

\subsection{Geometric interpretation of the momentum $\mu$}

We are going to prove that the quantity $\mu$ may be interpreted
as a hyperbolic angle between the vector normal to the external
Schwarzschild surface  $\{ t_{\textrm{Schw}} = \mbox{const.} \}$
and the vector normal to the internal Minkowski surface  $\{
t_{\textrm{Mink}} = \mbox{const.} \}$. The angle $\alpha ({\bf u},
{\bf v} )$ between the two normalized vectors ${\bf u}, {\bf v}$ is
defined by their (hyperbolic) scalar product:
\begin{equation}
\cosh \alpha ({\bf u}, {\bf v} ) := ({\bf u} | {\bf v}) \ .
\end{equation}

Similarly as in the Euclidean geometry, we call this quantity the {\em
angle between the two surfaces}: the Schwarzschild one and the
Minkowski one. To prove this interpretation of $\mu$ it is
sufficient to use formula (\ref{l-prime}). On the internal side of
the shell we use the second part of the formula and put $l = R^2$.
We then obtain
\begin{equation}\label{def-U}
\frac {l^\prime}{2 \sqrt{l}n} = \sqrt{ 1 + U^2 } \ .
\end{equation}
But inside of the shell the geometry  of ${\cal C}$ is given by a
three-dimensional spherically symmetric surface in the Minkowski
space. It is a matter of  straightforward calculations to prove
that in the Minkowski space-time the quantity on the left hand side of
the above equation is equal to $\cosh \alpha$, where $\alpha$ is
precisely the angle between such a subspace and the Minkowski flat
surface $\{ t_{\textrm{Mink}} = \mbox{const.} \}$. This implies
that  $U = \sinh \alpha$. But our surface is a smooth extension of
the external side of the Schwarzschild space $\{ t_{\mbox{\rm Schw}} =
\mbox{const.} \}$. This finally proves that $\mu = \alpha$ is the
angle between the leaves of the three-dimensional foliations of the
Schwarzschild and  the Minkowski spaces.

\subsection{Reduction of material part of the symplectic
form}\label{mat-reduction}

Let us first consider a simple non-trivial Lagrangian  of the
spinorial (Dirac) type (see  \ref{lagrangians} for the
discussion of the properties of null matter):
 \[
L=\sqrt{ \det g_{AB}}(z^2\dot{z}^1-z^1\dot{z}^2) \ .
\]
It implies the {\em second type}  constraints:
\[
P_1 = z^2 \ , \ \ \  P_2 = - z^1 \ ,
\]
which inserted into the symplectic form  (\ref{symplektyczna})
reduce it to the following form:
\begin{equation}\label{zredukowana}
  \Omega =  \frac 12 \delta \mu \wedge \delta \rho
  +8 \pi \delta (\sqrt{\rho} z^2 )\wedge \delta (\sqrt{\rho} z^1)
  \ .
\end{equation}
Moreover, denoting
\begin{eqnarray}\label{pq-red}
  p &:=&   \sqrt{8 \pi \rho} \ z^2 \ , \\
  q &:=&  \sqrt{8 \pi \rho} \ z^1 \ ,
\end{eqnarray}
leads to the following symplectic structure
\begin{equation}
\Omega=\frac 12 \delta \mu \wedge
\delta\rho\ + \delta p\wedge \delta q \ ,
\end{equation}
in a four-dimensional phase space parameterized  by variables:
$(p,q,\mu,\rho)$. Because the Hamiltonian (\ref{H-reduced}) does not
depend on the first pair of variables, hence their evolution is
trivial: they remain constant in time. And the evolution equations for
"geometric variables"  $(\mu ,\rho)$ generated by the Hamiltonian
(\ref{H-reduced}) are of the following form:
\begin{eqnarray}
\dot{\rho}&=&2\sqrt{\rho}e^{-2\mu}\label{r-ruchu11}\ ,\\
\dot{\mu}&=&-\frac12\frac{1}{\sqrt{\rho}}\left\{ 1 - e^{-2\mu}
\right\}\label{r-ruchu12}\ ,
\end{eqnarray}
the same  as in the case of the gravitational shock wave described in
the previous section.

It  turns out that a similar reduction is valid for the generic 
Lagrangian dependent on the two degrees of freedom (\ref{lagr-mat}), but, typically, the momentum $\mu$ canonically
conjugated to $\rho$ shall be modified by a function
$\sigma(p,q)$ giving the new momentum $\nu:= \mu - \sigma(p,q)$.
Consequently, the Hamiltonian (\ref{H-reduced}) depends upon the
variables $p$ and  $q$  via  $\mu = \nu + \sigma(p,q)$ and the
dynamics of the material variables is no longer trivial. We can
see that after performing the reduction of the material part of the symplectic
form (\ref{symplektyczna}) for the two degrees of freedom $K=1,2$, it
has the following form
\begin{equation}
\Omega=\frac 12 \delta  \mu \wedge \delta \rho +4 \pi
\delta (\rho P_K) \wedge\delta z^K\ ,
\end{equation}
From   (\ref{lagr-mat}) we have that:
\begin{equation}
P_K= F_K(z^1, z^2)\ ,
\end{equation}
hence
\begin{equation}\label{omega-int}
\Omega= 4\pi\delta \rho \wedge F_K \delta z^K+ 4 \pi\rho
F_{K,L}\delta z^L \wedge \delta z^K+\frac 12 \delta \mu \wedge
\delta \rho \ .
\end{equation}

The assumption  $dF \neq 0$ implies that $F_{2,1}- F_{1,2}\neq 0$.
It is then either positive or negative, and the form
$F_{K,L}\delta z^L \wedge \delta z^K$ is a symplectic form in two
dimensions. Besides, the Darboux theorem implies that there exists
such a coordinate system $\xi = \xi (z^1,z^2)$ and
$\eta=\eta(z^1,z^2)$ in which
\begin{equation}
F_{K,L}\delta z^L \wedge \delta z^K = \delta(F_K\delta
z^K)=\frac1{8\pi} \delta\left(\xi\delta\eta-\eta\delta \xi\right)
\ .
\end{equation}
Therefore,
\begin{equation}
F_K\delta z^K=\frac1{8\pi}\left(\xi\delta\eta-\eta\delta
\xi\right)+\frac1{8\pi}\delta\sigma\ ,
\end{equation}
where $\zeta=\sigma(\xi,\eta)$ is a function of variables
$(\xi,\eta)$. Inserting the above two equations to the formula
(\ref{omega-int}) we thus obtain $\Omega$ in the following form:
\begin{equation}
\Omega= \delta(\sqrt{\rho}\xi) \wedge \delta (\sqrt{\rho}
\eta)+\frac12\delta(\mu -\sigma)\wedge \delta\rho\ .
\end{equation}
We also introduce the new canonical variables
\begin{eqnarray}
p:&=&\sqrt{\rho}\ \xi\ ,\\ q:&=&\sqrt{\rho}\ \eta\ ,\\
\nu:&=&\mu-\sigma\
\end{eqnarray}
in order to simplify the form  $\Omega$:
\begin{equation}
\Omega=\delta p\wedge \delta q+\frac 12 \delta \nu\wedge
\delta\rho\ .
\end{equation}
The Hamiltonian of the system, still equal to the same expression
(\ref{H-reduced}), can be expressed in terms of the new variables as
follows:
\begin{equation} \label{H-reduced1}
 H ( \nu , \rho, p, q ) =
\frac12\sqrt{\rho}\left(1-e^{-2(\nu+\sigma\left(\frac{p}{\sqrt{\rho}},
\frac{q}{\sqrt{\rho}} \right))}\right) \ .
\end{equation}
This is a universal form of a Hamiltonian for a spherically symmetric
self-gravitating shell of null matter coupled to   two matter
fields. The properties of any specific model of such  matter are
uniquely implied by the function $\sigma$, dependent upon two
material variables $\left( \frac{p}{\sqrt{\rho}},
\frac{q}{\sqrt{\rho}} \right)$. This function is uniquely determined
by the two functions $F_K$ contained in the Lagrangian.

{\bf Example}: Consider the following Lagrangian density (see  \ref{lagr-ex})
$$ L=\sqrt{\det g_{AB}} \ \left(
(z^2+2z^1z^2)\dot{z}^1+((z^1)^2-z^1)\dot{z}^2\right)\ .
 $$
Rewriting this Lagrangian in the new variables $\xi$ and $\eta$ such
that $z^1=\frac12(\xi+\eta)$ and $z^2=\frac12 (\xi-\eta)$ we
obtain
$$ \frac12\sigma=(z^1)^2z^2=\frac18 (\xi+\eta)(\xi^2-\eta^2)\ .$$
The above equation expressed in terms of the canonical variables
$p=\sqrt{\rho}\xi=\sqrt{\rho}(z^1+z^2)$,
$q=\sqrt{\rho}\eta=\sqrt{\rho}(z^1-z^2)$ and $\rho$ takes the
following form: $$ \frac12\sigma=\frac18
\rho^{-3/2}(p+q)(p^2-q^2)\ .$$ 
The Hamiltonian of the whole system
reduces to:
$$
 H ( \nu, \rho, p, q ) =
\frac12\sqrt{\rho}\left(1-e^{-2\mu- \frac1{2\rho^{3/2}}
(p+q)(p^2-q^2)}\right) . $$

\section{Dynamics. Reconstruction of space-time
geometry}\label{dynamics}

The Hamiltonian  (\ref{H-reduced1}) uniquely   generates  the dynamics of
our system in the form of the following the Hamilton equations for the
canonical variables $(\rho, \nu )$ and $(p,q)$:
\begin{eqnarray}
\frac12 \dot{\rho}&=&\frac{\partial H}{\partial
\nu}=\sqrt{\rho}e^{-2(\nu+\sigma)}  \ ,\label{r-ha1}\\
\frac12\dot{\nu}&=&-\frac{\partial H}{\partial
\rho}=-\frac14\frac1{\sqrt{\rho}}(1-e^{-2(\nu+\sigma)})
-\sqrt{\rho}e^{-2(\nu+\sigma)}\frac{\partial \sigma(\rho,
p,q)}{\partial \rho}  \ , \label{r-ha2}\\
\dot{q}&=&\frac{\partial H}{\partial
p}=\sqrt{\rho}e^{-2(\nu+\sigma)}\frac{\partial
\sigma(\rho, p,q)}{\partial p}\ ,\label{r-ha3}\\
\dot{p}&=&-\frac{\partial H}{\partial
q}=-\sqrt{\rho}e^{-2(\nu+\sigma)}\frac{\partial \sigma(\rho,
p,q)}{\partial q}\ .\label{r-ha4}
\end{eqnarray}
The first two equations can be written in term of $\mu=\nu+\sigma$
and $\rho$ only. Hence  (\ref{r-ha1}) takes the  universal
form, identical to (\ref{r-ruchu11}):
\begin{equation}\label{r-ham1}
\dot{\rho}=2\sqrt{\rho}e^{-2\mu}\ .
\end{equation}
Combining it  with   (\ref{r-ha2}), (\ref{r-ha3}) and (\ref{r-ha4}) leads to:
\begin{equation}\label{dot-alpha}
\dot{\sigma}=-\frac1{\rho}e^{-2(\nu+\sigma)}\left( p
\frac{\partial\sigma}{\partial(\frac{p}{\sqrt{\rho}})}
+q\frac{\partial\sigma}{\partial(\frac{p}{\sqrt{\rho}})}\right)= 2\sqrt{\rho}e^{-2(\nu+\sigma)}
\frac{\partial\sigma}{\partial\rho}\ .
\end{equation}

Again,   (\ref{r-ha2}) and (\ref{dot-alpha}) imply the
universal equation, identical to  (\ref{r-ruchu12}):
\begin{equation}\label{r-ham2}
\dot{\mu}=-\frac12\frac1{\sqrt{\rho}}\left(1-e^{-2\mu}\right)\ .
\end{equation}
We conclude that the dynamics of the ``geometrical'' variables $(\rho ,\mu
)$ does not depend upon the choice of a model of matter and in
the case $\mu = \nu$ is identical to the one from the above Example. The
dynamics is also identical to the case of a gravitational shock wave
in empty space-time with no matter.

Once we know the dynamics of variables $(\rho ,\mu )$ we can
uniquely (up to the gauge) reconstruct the space-time in which
the dynamics is realized. Suppose that we have an explicit
solution of  (\ref{r-ham1}) and (\ref{r-ham2}). Choose
a gauge $\beta < -1$ and, separately for each moment of time,
gauge variables  $\zeta$ and  $l$. This enables us to entirely reconstruct
the set of the Cauchy data at each instant of time
separately. To reconstruct the whole geometry of space-time we also
need  lapse and shift functions. For this purpose let
us write the Einstein equations in terms of canonical variables
$g_{kl}$ and $P^{kl}$. Because those objects are already known (as
well as their time derivatives) at each moment of time, in result
we obtain elliptic equations for the lapse and  the shift. In this
way  for the lapse function  we obtain the second order equation in
variable  $r$ as a condition for preserving of the $\beta$-gauge
in time. This equation has to be solved with the following
boundary conditions:  $N = 1$ at infinity and $\frac {dN}{dr} = 0$
at $r=0$. In order to calculate the shift function we have to use
the equation for a time derivative of a three-dimensional metric.
This is an equation of the first order with respect to the shift
function and enables us to reconstruct it uniquely.

\subsection{Solution of the Hamilton equations}\label{ham-eq}

 From (\ref{r-ham1}) and
(\ref{r-ham2}):
\begin{eqnarray}
\dot{\rho}&=&2\sqrt{\rho}e^{-2\mu}\label{r-ruchu1}\ ,\\
\dot{\mu}&=&-\frac12\frac{1}{\sqrt{\rho}}\left\{ 1 - e^{-2\mu}
\right\}\label{r-ruchu2}\ ,
\end{eqnarray}
there follows
\begin{equation}
\frac{d\rho}{d\mu}=-4\rho\frac{e^{-2\mu}}{1-e^{-2\mu}}\ .
\end{equation}
Solution of the above equation is of the  form
\begin{equation}
\sqrt{\frac{\rho}{\rho_0}}=\frac{1}{1-e^{-2\mu}}\ .
\end{equation}
Therefore,  we have the  equation for the time evolution
of
 $\mu$:
\begin{equation}
\dot{\mu}=-\frac12
\frac1{\sqrt{\rho_0}}\left(1-e^{-2\mu}\right)^2\ ,
\end{equation}
whose solution reads as
\begin{equation}\label{rozw-mu}
\log(e^{2\mu}-1)-\frac{1}{e^{2\mu}-1}=-\frac1{\sqrt{\rho_0}}(t-t_0)\
.
\end{equation}
The constant  $\rho_0$ may be expressed in terms of the whole
energy of the system: $\sqrt{\rho_0}=2E$. The solution of 
(\ref{r-ruchu1}) for the time evolution of  $\rho$ is as follows
\begin{equation}
\sqrt{\rho}-E\log\rho= t-t_0\ .
\end{equation}

\section{Transition from the Schwarzschild time to the Minkowski time }
\label{time}

The equations (\ref{r-ham1}) and (\ref{r-ham2}) are the same as
the Hamilton equations for variables  $(\mu,\rho)$ in case when
$\sigma=0$. Then, $\nu=\mu$ and $(\mu,\rho)$ are canonical
variables. Because of the identical form of the dynamical equations,
in further calculations we limit ourselves to this particular
case.

Up to now, we have described the evolution of our system with respect
to the Schwarzschild time which coincides with the Minkowski time
at the spatial infinity. We can also describe our evolution with
respect to any other time variable, corresponding to a different
(3+1)-foliation of space-time. The leaves $S^{\cal G}_t$ of the new
foliation do not have to coincide with the previous hypersurfaces
$S_t=\{t=\textrm{const.}\},$ but we assume that asymptotically (at
space-infinity) they do.

A change of the time variable is not a standard transformation in
classical mechanics. Here, we use the approach introduced by one of
us in  \cite{shell1b}, based on the notion of a contact
manifold.

Suppose that the leaves $S^{\cal G}_t$ of the new foliation
intersect the shell at the Schwarzschild time $t+v(\rho,\mu)$. The
quantity $v$ will be called a {\em retardation} of the new time
variable with respect to the previous one, calculated on the
shell. Suppose that the value of this retardation depends upon the
actual dynamical situation, i.e. upon the position and the momentum, but
it does not depend explicitly upon the time variable. This means
that the new gauge condition is intrinsic, depending only on
initial data. The function $v=v(\rho,\mu)$ contains the entire
information about the transition between the old variables
$(\rho(t),\mu(t))$ and the new ones $(\rho^{\cal G} (t),\mu^{\cal
G}(t))$ because once we know $(\rho(t),\mu(t))$ we can solve the
equations generating dynamics of a system and set
\begin{eqnarray}
\rho^{\cal G}(t):&=& \rho\left(t+v(\rho(t),\mu(t))\right)\ ,\\
\mu^{\cal G}(t):&=& \mu\left(t+v(\rho(t),\mu(t))\right)\ .
\end{eqnarray}
For an arbitrary, not constant, function
 $v$, such a transformation is,  in general, not canonical. We will
 show in the sequel how to specify the canonical structure
of our reduced phase space in terms of these new variables. For
that purpose it will be convenient to use the language of contact
manifolds. Observe that the entire information about the dynamics of a
system may be retrieved from the three-dimensional contact space,
defined as the surface $\{E=H(\rho,\mu)\}$ in the four-dimensional
space $\{t, E,\rho,\mu\}$ equipped with the standard contact form:
\begin{equation}
\Psi:=\frac12 \delta \mu\wedge\delta \rho-\delta E\wedge \delta t
\ .
\end{equation}
This symplectic form in the four-dimensional phase-space becomes
degenerate when restricted to the surface $\{E=H(\rho,\mu)\}$.
Trajectories of the system are defined uniquely as those whose
tangent vector belongs to this degeneracy. To prove this it is
sufficient to parameterize our subspace in terms of three
variables $(t,\rho,\mu)$, and rewrite the form in the following
way:
\begin{equation}
\Psi:=\frac12 \delta \mu\wedge\delta \rho-\left(\frac{\partial
H}{\partial \mu}\delta \mu+ \frac{\partial H}{\partial
\rho}\delta\rho\right) \wedge \delta t\ .
\end{equation}
We see that the vector annihilating the above form must be
proportional to the vector
\begin{equation}
Z:=\frac{\partial}{\partial t}+\dot{\rho}\frac{\partial}{\partial
\rho}+\dot{\mu}\frac{\partial}{\partial \mu}\ ,
\end{equation}
where $\dot{\rho}$ and $\dot{\mu}$ are given by the Hamilton equations
(\ref{r-ham1}) and (\ref{r-ham2}).

In  \cite{shell1b} it was shown that the choice of the new
gauge condition ${\cal G}$ is equivalent to the choice of the variable $T$
\begin{equation}\label{def-te}
T:=t-v
\end{equation}
as a new time variable. Let us rewrite, therefore, our symplectic
form $\Psi$ in terms of the new time  $T$. It will be convenient
to have the energy $E$ instead of $\mu$ as an independent parameter and
treat   $\mu$ as a function  $\mu=\mu(E,\rho)$ obtained from
solving  (\ref{H-reduced}). Therefore, we have 
\begin{eqnarray}
\Psi&=&\frac 12 \delta\mu\wedge\delta\rho-\delta E\wedge\delta T -\delta
E\wedge\left(\frac{\partial v}{\partial E}\delta E+\frac{\partial
v}{\partial\rho}\delta\rho\right)\nonumber\\&=&
\frac12\left(\delta\mu-2\frac{\partial v}{\partial \rho}\delta
\rho\right)\wedge\delta\rho-\delta E\wedge\delta T\ .
\end{eqnarray}
Now define:
\begin{equation}
V(E,\rho):=2\int \frac{\partial v}{\partial \rho}(E,\rho)d
\rho+a(\rho)=\int \frac1R  \frac{\partial v}{\partial R}(E,R)d
\rho+a(\rho)\ ,
\end{equation}
where $a(\rho)$ is arbitrary. Then, we have
\begin{equation}
  \Psi=\frac12\delta\tilde{\mu}\wedge\delta\rho-\delta E\wedge\delta
  T\ .
\end{equation}
where
\begin{equation}\label{def-tilde-mu}
\tilde{\mu}:=\mu - V\left(H(\rho,\mu),\rho\right)
\end{equation}
plays the role of the momentum canonically conjugate to $\rho^{\cal
G}(t)$. Differentiating  (\ref{def-tilde-mu}) with respect to  $E$
provides
\begin{equation}\label{diff-tilde}
\frac{\partial}{\partial E}(\tilde{\mu}-\mu)=-2\frac{\partial
v}{\partial\rho}(E,\rho)=-\frac1R\frac{\partial v}{\partial
R}(E,R)\ .
\end{equation}

It is interesting to notice (cf. \cite{shell1b}) that any choice
of the function $\tilde{\mu}= \tilde{\mu} (\mu,\rho)$ leads to a
certain gauge condition, i.e. to the definition of a new time
parameter on the shell because we may always reconstruct the
retardation function $v$ from (\ref{diff-tilde}). As an example,
consider $\tilde{\mu}=U n(\zeta)=4\pi p/R $ (see the formula
(\ref{U})) which corresponds to the Minkowski time calculated at
the internal side of the shell. Indeed, in the Minkowski spacetime the
momentum canonically conjugate to the shell's position $R$ is
equal to the kinetic momentum $p$. The change of the variables from $R$ to
$\rho$ implies a transformation of momenta according to the identity:
\[
   p \ \textrm{d} R = \frac 12 \frac p{R} \textrm{d} \rho \ ,
\]
which, finally, gives $\frac 12 \tilde{\mu} \textrm{d} \rho$ when
integrated over the shell. We are going to show in the sequel
that the evolution defined in this way corresponds, indeed, to the
Minkowski time. For this purpose we express the new momentum in
terms of the old one using (\ref{mom-can}):
\begin{equation}
  \tilde{\mu}=\sinh\mu\sqrt{1-\frac{2E}{R}}= \sinh\mu\ e^{-\mu}
  = \frac 12 \left( 1 - e^{-2\mu} \right) \ ,\label{e-r}
\end{equation}
where the energy $E$ is equal to the value of the Hamiltonian $H$.
But, from (\ref{H-reduced}), we obtain:
\begin{equation}\label{mu-def}
\sqrt{1-\frac{2E}{R}}= e^{-\mu}\ ,
\end{equation}
and, consequently,
\begin{equation}
\tilde{\mu}=\frac{E}{R} \ ; \ \ \ \ \ \ \ \frac{\partial
\tilde{\mu}}{\partial E}=\frac{1}{R}\ .
\end{equation}
On the other hand, we get from (\ref{mu-def}):
\begin{equation}
\frac{\partial \mu}{\partial E}=\frac{1}{(R-2E)}\ .
\end{equation}
Thus,  (\ref{diff-tilde}) implies:
\begin{equation}\label{v-po-r}
\frac{\partial v}{\partial R}=\frac{2E }{R-2E}\ .
\end{equation}
To prove that  the above quantity describes retardation between
the Minkowski time  $T$ inside the shell an the extrinsic
Schwarzschild time $T$, differentiate  (\ref{def-te})
over $t$:
\begin{equation}
\dot{T}=1-\frac{\partial v}{\partial R}\dot{R}\ ,
\end{equation}
where we have used the energy conservation $\dot{E}\equiv 0$. The
derivative $\dot{R}$ can be calculated from 
(\ref{r-ham1}), and
 relation (\ref{mu-def}), leading to:
 \begin{equation}
\frac{\partial v}{\partial R}\dot{R}=\frac{2E}{R}
\end{equation}
and finally
\begin{equation}
\dot{T}=1-\frac{2E}{R}\ .
\end{equation}
Integrating  (\ref{v-po-r}) we obtain a retardation between the
Minkowski time $T$ and the Schwarzschild time $t$:
\begin{equation}
v(E,R)=2E\log(R-2E)\ ,
\end{equation}
hence
\begin{equation}\label{t-min-sch}
T=t+2E\log(R-2E)=t+2R\tilde{\mu}\log R\left(1-2\tilde{\mu}\right)\ ,
\end{equation}
and $R>2E$, otherwise $\mu$ would not be defined (Eq. \ref{e-r}). The
Hamiltonian may be expressed in terms of the Minkowski variables
$(\rho , \tilde{\mu})$:
\begin{equation}
H=\sqrt{\rho}\tilde{\mu}
\end{equation}
and   the Hamilton equations take the form:
\begin{eqnarray}
\frac12 \frac{d}{dT}\rho&=&\sqrt{\rho}\ , \label{r-ruchuM1}\\
\frac12 \frac{d}{dT}\tilde{\mu}&=&-\frac12
\frac{\tilde{\mu}}{\sqrt{\rho}}\ . \label{r-ruchuM2}
\end{eqnarray}
Using the relations  $\frac{d}{dT}= \frac{dt}{dT}\frac{d}{dt}$ and
(\ref{t-min-sch}) between $t$ and $T$ we can reconstruct 
(\ref{r-ruchu1}) and (\ref{r-ruchu2}).

The dynamical equations (\ref{r-ruchuM1}) and (\ref{r-ruchuM2}) may be
explicitly solved:
\begin{eqnarray}
\sqrt{\rho}&=&T-T_0 + \sqrt{\rho_0}\ ,\\
\tilde{\mu}&=&\tilde{\mu}_0\frac{\sqrt{\rho_0}}{T-T_0+\sqrt{\rho_0}}
\ .
\end{eqnarray}
Keeping in mind that $\sqrt{\rho}=R$, the above solution for $R$
(linear propagation with velocity equal to $1$) is implied by the fact
that the only null-like, spherically symmetric surfaces in the
Minkowski space are light-cones. Hence, the evolution cannot be
global, it ends at the cone's vertex. An interesting case of the {\em
global} (in the Minkowski time) evolution is obtained when the two
cones: the future oriented one and the past oriented one, cross
each other. The theory of crossing shells was thoroughly analyzed
in  \cite{sh-cross}.

\section{Conclusions }\label{conclusions}
We have derived from first principles the Hamiltonian dynamics of a spherically symmetric shell of null matter. For
this purpose we have imposed a strong condition: continuity of
all 10 components of the metric across the shell. This is {\em
not} a physical restriction, but only a gauge condition imposed on
possible coordinate systems which can be used. Due to this gauge
condition we were able to define a singular (Dirac-delta like)
Riemann tensor of a non-continuous spacetime connection $\Gamma$
in terms of the standard formulae of differential geometry, where
derivatives are understood in the sense of distributions. Imposing
spherical symmetry leads to an effective reduction of the phase
space of the system. The resulting ``reduced phase space'' is
obtained as a quotient of an {\em infinite-dimensional} space of
 Cauchy data with respect to the (again {\em infinite-dimensional})
group of gauge transformations generated by the Gauss-Codazzi
constraints. Such a complete reduction which, miraculously, leads
to a {\em finite dimensional} phase space, has not been
successfully performed up to now. In particular, the radial
component of the constraint equations produces the ``equation of
state'' (\ref{eqos}): a relation between the radial component of the
momentum $p$ and the surface energy density $\epsilon$ on $S$. In
our approach the relation is not postulated but is derived as a
consequence of the nullness of matter.

Choosing $\rho=R^2$ (where $R$ is the physical radius of the
shell) as a configuration parameter of the shell, the
corresponding canonical momentum was shown to be the (hyperbolic)
angle $\mu$ between the two foliations: the Schwarzschild foliation
outside of the shell and the Minkowski foliation inside of it. The
ADM energy calculated at space infinity, expressed in terms of the
canonical variables $(\rho, \mu)$ was proved to be the Hamiltonian
of the complete ``gravity + matter'' system. It generates the
dynamics of the canonical variables and  enables us to uniquely reconstruct
(up to an arbitrary choice of the gauge) the spacetime
dynamics.

We have also discussed the dependence of the above picture upon
the choice of the time variable. To implement such a
transformation at the level of the Hamiltonian dynamics, we have
proposed a new method based on the contact structure of the
enlarged phase space. To illustrate this method we have shown how
to transform the entire theory from the Schwarzschild time, measured
at space infinity, to the Minkowski time, measured inside the
shell.

\appendix
\section{Null matter }\label{lagrangians}
Because the shell metric $g_{ab}$, $a,b=0,1,2;$ is degenerate, we
have {\em a priori} no standard scalar density on $S$ which can be
used in the definition of the Lagrangian. We must, therefore,
manufacture such a density from the following ingredients: i)
the matter fields $z^{K}$, ii) their derivatives $z^{K}_{,a}$ along
the shell, iii) the degeneracy field $X$ of $g_{ab}$ on $S$ and
iv) the two-dimensional volume form $\lambda:= \sqrt{\det g_{AB}}$
on each surface $\{x^0 = \mbox{\rm const.}\}$. Choose the field
$X$ (otherwise given up to a multiplicative constant) in such a
way that $< dx^0 ; X> = 1$. The following object: $L=\sum_K
\lambda X^a z^{K}_{,a}$, ($K=1\dots n$), is a scalar density and
may be taken as a matter Lagrangian. A more sophisticated example
is given by $ L=\lambda\left(X^a z^K_{\;,a}X^b
z^L_{\;,b}\sigma_{KL}(z^A)\right)^\alpha
\left(X^c\xi_{,c}\right)^{\beta}$, where $\xi$ is a scalar field,
$2\alpha+\beta=1$ and $\sigma_{KL}$ denotes any metric tensor in
the space of field variables. However, in the case of spherical symmetry we have $X^a =
\delta^a_0$ and all these models lead to the trivial Hamiltonian
$H=0$, which implies the equation $\dot{\lambda}=0 $. The shell
surface must be, therefore, an isolated horizon (cf. \cite{JKC},
Eq. (6.2))

We are interested in examples of null-matter which may non-trivially couple
to a generic null shell, not necessarily an isolated
horizon. Such a nontrivial example can be obtained if one
considers a thin shell of matter described by the Lagrangian
density with at least two degrees of freedom which depends not
only on velocities, but also on the configuration variables.
Consider the Lagrangian density of the type:
\begin{equation}
  \label{lagrangian-main}L=L( z^K ; z_{,a}^K)=\lambda X^az_{,a}^K F_K(z^L) \ ,
\end{equation}
where $K,L=2\dots n$, and  $F_K$ is a covector field defined on
the space of material variables. We show in a sequel that this
leads to a non-trivial model already for the two degrees of freedom:
$K,L=1,2$. In a spherically symmetric case this Lagrangian takes the
form:
\begin{equation}\label{lagr-mat}
L=\lambda \dot{z}^K F_K(z^L) \ .
\end{equation}
The Euler-Lagrange equations for this system read as
\begin{eqnarray}
\dot{\lambda}F_1&=\lambda(F_{2,1}-F_{1,2})\dot{z}^2\ ,\\
\dot{\lambda}F_2&=-\lambda(F_{2,1}-F_{1,2})\dot{z}^1\ .
\end{eqnarray}
If the quantity  $(F_{2,1}-F_{1,2})$ is equal to zero, i.e. if $F$
is closed: $dF=0$, then $\dot{\lambda}=0$, and matter described by
this Lagrangian density couples again to an isolated horizon.
Assume that $dF\neq 0$. Then the Euler-Lagrange equations imply
the following constraints equations:
$$F_1\dot{z}^1 + F_2\dot{z}^2 = 0\ .$$
\hfill$\Box$
\section{Examples of non-trivial matter Lagrangians}\label{lagr-ex}
1. A simple example of a Lagrangian of the type
(\ref{lagrangian-main}) may be the following one:
\begin{equation}\label{dirac}
L=\lambda(z^2\dot{z}^1-z^1\dot{z}^2) \ ,
\end{equation}
whose   structure resembles the Dirac Lagrangian for spinor
fields. This Lagrangian implies the following constraints:
\begin{eqnarray}
  \lambda P_1  &=& p_1 = \frac {\partial L}{\partial \dot{z}^1}= \lambda
  z^2  \\
  \lambda P_2 &=& p_2 = \frac {\partial L}{\partial \dot{z}^2}= - \lambda
  z^1
\end{eqnarray}
These are {\em second type} constraints. Inserting them into the
symplectic form  (\ref{symplektyczna}) we obtain:
\begin{equation}\label{zredukowana-1}
  \Omega =  \frac 12 \delta \mu \wedge \delta \rho
  +8 \pi \delta (\sqrt{\rho} z^2 )\wedge \delta (\sqrt{\rho} z^1)
  \ .
\end{equation}
Similarly as in the spinor theory, one  configuration variable (in
electrodynamics it is the imaginary part of spinor in the Majorana
representation) becomes  momentum canonically conjugated to the
second variable (the real part in the Majorana representation).

2. Another example:
\begin{equation} L=\lambda\left(
(z^2+2z^1z^2)\dot{z}^1+((z^1)^2-z^1)\dot{z}^2\right)\ .
\end{equation}
is used to reduce the symplectic form described in Section
\ref{mat-reduction}.

\section*{References}

\end{document}